\documentstyle[11pt,a4]{article}
\begin{document}
\begin{center}
{\large {\bf Erratum for "Nucleosynthesis Constraints on Active-Sterile...}}" \\
{\large by}\\
{\large \bf  V. B. Semikoz and J. W. F. Valle}, \\
{\large  published in }\\
{\sl Nucl. Phys. } {\bf B425} (1994) 65.
\end{center}
\vskip .2cm
\def\lsim{\raise0.3ex\hbox{$\;<$\kern-0.75em\raise-1.1ex\hbox{$\sim\;$}}}
\def\gsim{\raise0.3ex\hbox{$\;>$\kern-0.75em\raise-1.1ex\hbox{$\sim\;$}}}
\def\VEV#1{\left\langle #1\right\rangle}
\let\vev\VEV
\def\eq#1{{eq. (\ref{#1})}}
\def\apj#1#2#3{          {\it Astrophys. J. }{\bf #1} (19#2) #3}
\def\app#1#2#3{          {\it Astropart. Phys. }{\bf #1} (19#2) #3}
\def\ib#1#2#3{           {\it ibid. }{\bf #1} (19#2) #3}
\def\nat#1#2#3{          {\it Nature }{\bf #1} (19#2) #3}
\def\nps#1#2#3{        {\it Nucl. Phys. B (Proc. Suppl.) }{\bf #1} (19#2) #3} 
\def\np#1#2#3{           {\it Nucl. Phys. }{\bf #1} (19#2) #3}
\def\pl#1#2#3{           {\it Phys. Lett. }{\bf #1} (19#2) #3}
\def\pr#1#2#3{           {\it Phys. Rev. }{\bf #1} (19#2) #3}
\def\prep#1#2#3{         {\it Phys. Rep. }{\bf #1} (19#2) #3}
\def\prl#1#2#3{          {\it Phys. Rev. Lett. }{\bf #1} (19#2) #3}
\def\ppnp#1#2#3{           {\it Prog. Part. Nucl. Phys. }{\bf #1} (19#2) #3}
\def\tp{these proceedings}
\def\pc{private communication}
\def\opc{\hbox{{\sl op. cit.} }}
\def\ip{in preparation}

The magnetization asymmetry $M_j^{(a)}-M_j^{(\tilde{a})}$ 
given in eq. 3.6 of ref. \cite{SVNPB} has a wrong relative
sign between particle and anti-particle contributions. 
Here we present the derivation of the correct sign
and its implications for the limits derived in ref. 
\cite{SVNPB}.  We start from eq. (2.9) of \cite{SVNPB}, 
\begin{equation}
\label{n1}
M_j^{(a)}-M_j^{(\tilde{a})} \equiv 
\mu_B
\vev{\bar{\Psi}_a \gamma_j \gamma_5 \Psi_a}_0 =
\mu_B \vev{\Psi_a^\dagger \Sigma_j \Psi_a}_0 
\end{equation}
where $\vev{}_0$ is an averaged axial-vector current of the $a$-type 
particle in the medium, with a magnetic field {\bf B}=(0,0,$B$) along 
the z-axis. The Fermi distributions for the case of leptons are given by 
\begin{equation}
\label{n2}
\vev{ b_\lambda^{a\dagger} b_{\lambda'}^a}_0=
f^{(a)}_{\lambda' \lambda}(k_z,n) =
\frac{\delta_{\lambda \lambda'}}
{\exp[(\epsilon_{n\lambda}(k_z)-\zeta_a)/T]+1}
\end{equation}
where the spectrum is 
\begin{equation}
\label{n3}
\epsilon_{n\lambda}(k_z) = \sqrt{k_z^2+m_a^2+\mid e \mid B
(2n+1) - | e| \lambda B}
\end{equation}
Here $k$ is the momentum, $n$ the Landau number ($n=0,1,2,\ldots$) 
and $\lambda$ is twice the eigenvalue of the conserved spin projection 
on the magnetic field, 
$(\sigma_z)_{\lambda \lambda'}= \lambda \delta_{\lambda \lambda'}$.
For anti-leptons one has
\begin{equation}
\label{n4}
\vev{ d_\lambda^{a\dagger} d_{\lambda'}^a}_0=
f^{(\tilde{a})}_{-\lambda, -\lambda'}(k_z,n)=
\frac{\delta_{-\lambda,-\lambda'}}
{\exp[(\tilde{\epsilon}_{n\lambda}(k_z)+\zeta_a)/T]+1}
\end{equation}
where the change in sign of the last term in the spectrum accounts 
for the opposite electric charge,
\begin{equation}
\label{n5}
\tilde{\epsilon}_{n\lambda}(k_z) = \sqrt{k_z^2+m_a^2+\mid e \mid B
(2n+1) + | e| \lambda B}
\end{equation}
The lepton contribution $M_j^{(a)}$ to the magnetization asymmetry 
defined in \eq{n1} is correctly given by eq. (3.3) of \cite{SVNPB} 
where, due to the degeneracy of states $\epsilon_{n,-1}=\epsilon_{n+1,1}$, 
only the $\epsilon_{0,1}$ term survives in summing over Landau numbers. 
In contrast, for anti-leptons one obtains a negative sign from the 
re-ordering of the anti-particle Fock operators $d^a$ and $d^{a\dagger}$ 
in the distribution function \eq{n4}. Note that this (-) sign {\sl has 
already been extracted} in the definition of the magnetization asymmetry,  
\eq{n1}. From the relation $v_a=C\bar{u}^T_a$ and the conjugation property  
$-\sigma_3^T=\sigma_2 \sigma_3 \sigma_2$ we can write
\begin{equation}
\label{n9}
- M_z^{(\tilde{a})} = - \mu_B \sum_{n=0}^{\infty} \frac{|e|B}{(2\pi)^2}
\int_{-\infty}^\infty dk_z~ Tr[-\sigma_3^T f^{(\tilde{a})}(k_z,n)]
\end{equation}
where
\begin{equation}
\label{n10}
Tr[-\sigma_3 f^{(\tilde{a})}] =
\sum_\lambda (-\lambda) f^{(\tilde{a})}_{-\lambda,-\lambda}
\end{equation}
Again the sum over Landau numbers is greatly simplified, due to
the degeneracy $\tilde{\epsilon}_{n,1}=\tilde{\epsilon}_{n+1,-1}$
that follows from \eq{n5}. The only term that contributes is that 
for which $n=0$ and $\lambda = -1$, leading to
\begin{eqnarray}
\sum_{n=0}^{\infty} \Bigl [ 
-\frac{1}{\exp[(\tilde{\epsilon}_{n,1}+\zeta_a)/T]+1} 
+ \frac{1}{\exp[(\tilde{\epsilon}_{n,-1}+\zeta_a)/T]+1}\Bigr ]
\nonumber \\
= \frac{1}{\exp[(\sqrt{k_z^2+m_a^2}+\zeta_a)/T]+1}
\label{n11}
\end{eqnarray}
As a result the magnetization asymmetry is obtained by {\sl subtracting} 
lepton and anti-lepton contributions, 
\begin{eqnarray}
\label{sign}
M_z^{(a)}-M_z^{(\tilde{a})} = \mu_B 
\frac{2|e|B}{(2\pi)^2}
\int_0^\infty dk_z~\Bigl [ 
\frac{1}{\exp[(\sqrt{k_z^2+m_a^2}-\zeta_a)/T]+1} 
-\frac{1}{\exp[(\sqrt{k_z^2+m_a^2}+\zeta_a)/T]+1}
\Bigr ]
\label{n12}
\end{eqnarray}
in contrast to our claim in \cite{SVNPB}. To summarize our derivation, 
notice that the relative minus in \eq{sign} follows from the three minus 
signs obtained above: one from operator ordering, one from the conjugation 
property, and one from the different way in which the Landau levels are 
summed over in the case of particles and anti-particles (different 
helicities in each case). This sign has also been recently derived 
in ref. \cite{elmfors}, using a different approach.  

As a result one should use for the magnetic field contribution 
$f(q,B)$ in Eq. (3.7) of \cite{SVNPB} the non-local form
\begin{equation}
\label{nonloc}
f(q,B) = 0.47 G_F e\vec{B}\vec{q} T^2/M_W^2   
\end{equation}
given by Eq. (12) of \cite{elmfors}. Note that our asymptotic 
solution for the active-sterile neutrino conversion probability 
Eq. (4.9) of [1] remains valid, as well as our final expression for 
that probability after averaging over fast neutrino collisions,  
Eq. (6.6) of [1]. However, the bounds on active-sterile neutrino 
mixing parameters derived for a hot charge-symmetric plasma 
derived in \cite{SVNPB} change substantially as a result of 
the new dispersion relation that follows from \eq{nonloc}.
Substituting \eq{nonloc} above into Eq. (6.6) of [1] (where 
$\vev{ \Delta_B^2}  \equiv \vev{f(q, B)^2} $) we can write the 
nucleosynthesis bound on neutrino mixing parameters as
\footnote{Here we assume that magnetic fields survive after
re-combination time and seed the galactic magnetic fields.},
\begin{equation}
\label{bound}
\mid \Delta m^2\mid \mid \sin 2\theta\mid \leq 0.76(2.75)^p\times
10^{- 12 - 8p}\Bigl ( \frac{T}{\mbox{MeV}}\Bigr )^{13/2 + 4p}, 
\end{equation}
where the r.m.s. field scaling parameter $p$ obeys $0\leq p\leq 3/2$.
For the case $p = 0.5$ analyzed in [1,2] the constraint in \eq{bound}
looks more restrictive than our Eq. (6.9) in [1]. However, the
self-consistent condition $\Gamma \gg  V_{\nu_e}$ [1] requires the
substitution of high temperatures $T \sim T_{QCD}\sim 150~\mbox{MeV}$ 
into \eq{nonloc} leading to
\begin{equation}
\label{newbound}
\mid \Delta m^2\mid \mid \sin 2\theta\mid \lsim 400~\mbox{eV}^2
\end{equation}
which is substantially weaker than Eq. (6.10) in [1]. This
revised constraint is more restrictive than the bounds obtained in 
the absence of random magnetic fields (Eq. (1.1) [1]) for very small
neutrino mixing, $\sin 2\theta \lsim 2 \times 10^{-3}$.

\vskip .5cm
We acknowledge Per Elmfors who sent us the paper \cite{elmfors}
before publication. We would like to stress that, unlike implied 
in \cite{elmfors}, the sign error in eq. (3.6) of \cite{SVNPB} 
does not affect in any way the conclusions reached in subsequent 
papers, such as those in ref. \cite{pastor}. We thank H. Nunokawa 
and S. Pastor for checking this derivation.

\end{document}